\documentclass[a4paper,aps,onecolumn,nofootinbib,preprintnumbers,superscriptaddress,amsmath,amssymb,amsfonts]{revtex4}
\usepackage{graphicx}
\usepackage{bm,natbib}
\usepackage{amsmath}
\usepackage[english]{babel}
\usepackage[T1]{fontenc}
\usepackage{amssymb}
\usepackage{amsfonts}
\usepackage{epsfig}
\usepackage{colordvi}
\usepackage{psfrag}
\usepackage{color}
\usepackage{ mathrsfs }

\newcommand{\Lagr}{\mathcal{L}}

\newcommand{\G}{\mathcal{G}}

\usepackage{dcolumn}
\usepackage{multirow}
\usepackage{hyperref}
\usepackage{epstopdf}
\usepackage{bm}
\usepackage{times}
\usepackage{comment}
\usepackage{ulem,lipsum}

\hypersetup{
  colorlinks=true,        
  linkcolor=blue,         
  citecolor=magenta,      
}

\begin{document}
\title{\textbf{Extended Theories of Electrodynamics in $f(R)$ Gravity}}


\author{Francesco Bajardi}
\email{f.bajardi@ssmeridionale.it}
\affiliation{Scuola Superiore Meridionale, Largo San Marcellino 10, I-80138, Naples, Italy.}
\affiliation{INFN Sez. di Napoli, Compl. Univ. di Monte S. Angelo, Edificio G, Via Cinthia, I-80126, Naples, Italy.}

\author{Micol Benetti}
\email{m.benetti@ssmeridionale.it}
\affiliation{Scuola Superiore Meridionale, Largo San Marcellino 10, I-80138, Naples, Italy.}
\affiliation{INFN Sez. di Napoli, Compl. Univ. di Monte S. Angelo, Edificio G, Via	Cinthia, I-80126, Naples, Italy.}

\author{Salvatore Capozziello}
\email{capozziello@na.infn.it}
\affiliation{Scuola Superiore Meridionale, Largo San Marcellino 10, I-80138, Naples, Italy.}
\affiliation{INFN Sez. di Napoli, Compl. Univ. di Monte S. Angelo, Edificio G, Via	Cinthia, I-80126, Naples, Italy.}
\affiliation{Dipartimento di Fisica ``E. Pancini", Universit\`a di Napoli ``Federico II", Complesso Universitario di Monte Sant’ Angelo, Edificio G, Via Cinthia, I-80126, Napoli, Italy,}

\author{Abedennour Dib}
\email{abedennour.dib@cnrs-orleans.fr}
\affiliation{\mbox{Laboratoire de Physique et Chimie de l'Environnement et de l'Espace (LPC2E) UMR 7328}\\
\mbox{Centre National de la Recherche Scientifique (CNRS), Universit\'e d'Orl\'eans (UO), Centre National d'\'Etudes Spatiales (CNES)}\\
\mbox {3A Avenue de la Recherche Scientifique, 45071 Orl\'eans, France}}

\affiliation{\mbox{Observatoire des Sciences de l'Univers en region Centre (OSUC) UMS 3116} \\
\mbox{ Universit\'e d'Orl\'eans (UO), Centre National de la Recherche Scientifique (CNRS)}
\mbox{ Observatoire de Paris (OP), Universit\'e Paris Sciences \& Lettres (PSL)}\\
\mbox{1A rue de la F\'{e}rollerie, 45071 Orl\'{e}ans, France}}
\affiliation{\mbox{D\'epartement de Physique, Unit\'e de Formation et Recherche Sciences et Techniques, Universit\'e d'Orl\'eans}\\
\mbox{Rue de Chartres, 45100 Orl\'{e}ans, France}}

\date{\today}

\begin{abstract}

Within the general framework of $f(R)$ gravity, we introduce a function of the electromagnetic curvature invariant $f(\mathbb{F})$ that couples minimally to gravitation to ensure a consistent treatment of curvature functions in these theories. We show that one of the solutions leads to field equations that are a generalization of the Klein-Gordon equation while the other leads to a typically non-linear massless solution. Focusing on flat spacetime, our formalism recovers the Plebanski family of models and Bopp-Podolsky electrodynamics as specific limits. These extensions may have phenomenological consequences in extreme environments, such as the early universe or near charged compact objects, where deviations from classical electrodynamics might be probed.
\end{abstract}


\maketitle

\section{Introduction}
The electromagnetic interaction is perhaps the most studied among the fundamental forces, due to its suitability at the quantum level and because it was the first to be well interpreted and supported by strong mathematical foundations. By construction, it describes massless particles with integer spin. The latter is a consequence of the vector nature of the electromagnetic field, while the absence of mass is an assumption motivated by experiments. In fact, the photon mass has never been experimentally detected and thus the entire theory is constructed in terms of massless particles. Since the nineteenth century, several experiments have confirmed the absence of photon mass and any  hypothetical photon mass should be extremely small if compared to other fundamental particles \cite{Bonetti:2016vrq}. The current upper limits on the photon mass $m_\gamma$ lie within the range $10^{-60} \text{kg} < m_\gamma < 10^{-50} \text{kg}$ \cite{ParticleDataGroup:2024cfk, Bonetti:2016cpo, Ryutov:1997zz, Ryutov:2007zz, Retino:2013gga, Goldhaber:1971mr,Spallicci:2022eoe}. Since the discovery of neutrinos oscillations, the photon has become the last remaining particle in the Standard Model that is both free and massless (gluons are massless, but are confined and gravitons are not included in the Standard Model). Investigations on the possibility of a very tiny mass, enlarging the theoretical structure of standard Electrodynamics, have been on-going \cite{Bonetti:2017pym, Spallicci:2022eoe, Heisenberg:2014rta, Podolsky:1944zz,Dib:2025qoy}.

Over the years, several electromagnetic theories have been developed in order to extend or modify the Maxwellian theory, introducing the possibility of photons with nonzero mass \cite{Proca:1936fbw, Spallicci:2022eoe, Bopp1940, Podolsky:1942zz, Heisenberg:2014rta}. 
Among these, the de Broglie-Proca (dBP) theory \cite{Proca:1936fbw} is one of the earliest linear extensions, where the electromagnetic field is described by a massive vector potential, leading to the breaking of gauge symmetry. While initially this might seem to rule out the gauge theoretical nature of the dBP model, it is possible to show that the model is nothing but the gauge fixed version of the Stückelberg model \cite{Ruegg:2003ps} where $\phi=0$. Another significant approach is provided by non-linear electromagnetic theories, such as the Born-Infeld theory \cite{Born:1934mzs}, which introduce nonlinear corrections to the electromagnetic Lagrangian, often motivated by quantum considerations or attempts to regularize field singularities. In some cases \cite{Spallicci:2024drl,Dib:2025qoy}, these theories predict modifications in the propagation of electromagnetic waves that can be interpreted as effects associated with massive photons or new polarization modes.

Other important formulations include higher-order theories, the first of which was the so-called Bopp-Podolsky theory \cite{Bopp1940, Podolsky:1942zz}, a linear extension of electrodynamics that modifies the action by introducing higher-derivative terms, which can be interpreted as implying the presence of an additional massive photon.

Extended theories of Electromagnetism are considered for various important reasons. Primarily, classical Maxwell theory, while extraordinarily successful, faces limitations when applied to extreme regimes such as very high energies, strong gravitational fields, or at quantum scales, where phenomena like vacuum polarization occur. Introducing modifications such as a non-zero photon mass or higher-derivative terms can potentially explain small deviations from classical predictions and provide effective descriptions of new physics \cite{Goldhaber:2008xy, Kostelecky:2008ts}. For instance, in strong gravitational backgrounds, extended electromagnetic theories may better capture interactions between electromagnetic and gravitational fields \cite{Podolsky:1942zz}. Massive photons could be suitable candidates for some dark matter models (Dark photons, Ultra light dark matter, fuzzy dark matter etc.) while simplifying the quantization procedure by getting rid of infrared divergences.

Although the present work involves a substantial formal development, its motivation and scope are primarily physical. The extended electromagnetic framework introduced here is designed to capture deviations from standard Maxwell theory that are expected to arise in extreme regimes, such as strong gravitational fields, high-energy environments, or cosmological settings where quantum or semiclassical corrections become relevant. The mathematical structure of the theory is therefore a necessary tool to consistently identify the physical degrees of freedom, their propagation properties, and the conditions under which massive photon-like modes can emerge without explicit symmetry breaking. In this sense, the derivation of the field equations and of the associated energy–momentum tensor provides the basis for addressing concrete physical questions related to photon propagation, conformal symmetry breaking, and effective mass generation in curved spacetime, with potential implications for early-universe physics and high-energy astrophysics.

Moreover, such extensions often appear naturally within the framework of effective field theories derived from more fundamental theories, such as string theory or quantum gravity approaches \cite{Polchinski:1998rq, Zwiebach:1985uq}. They can incorporate nonlinearities, additional polarization states, or new couplings that classical Electromagnetism cannot describe. Studying these theories therefore allows for probing physics beyond the Standard Model and deepening our understanding of fundamental interactions under extreme conditions \cite{Kostelecky:2008ts}.

In this work, we are going to consider a generic extension of electrodynamics which, under certain conditions, reduces to the Bopp-Podolsky model in curved space-time, implying the presence of a massive gauge boson due to higher-order terms in the electromagnetic Lagrangian and gravitational coupling. We start by introducing the most well-known massive electromagnetic theories, namely the Proca theory and the three-dimensional Chern-Simons theory. Then, we focus on a general extension of the electromagnetic action including a coupling with the gravitational field and other higher-order corrections to the electromagnetic tensor $F_{\mu \nu}$. Among all possible extensions, we focus on the one leading to Klein-Gordon–like equations and including a new massive vector with an additional polarization mode.

The layout of the paper is the following. In Sec.\ref{MTE}, we describe the general scheme of massive theories of Electromagnetism. Sec.\ref{EE} is devoted to the development of the Extended Electromagnetism. We discuss the results and draw our conclusions in Sec.\ref{Conc}.

Throughout the paper, we will use natural units: $\hbar = c = k_B = 8\pi G = e^+ = 1$.

\section{Massive Theories of Electromagnetism}\label{MTE}
In this section, we present two well-established models of massive electrodynamics: the Proca theory and the three-dimensional Chern-Simons theory \cite{Proca:1936fbw, Deser:1982vy}. We begin by introducing the standard action for free, massless electrodynamics:
\begin{equation}
S = \int \text{d} \textbf{A} \text{d} \textbf{A},
\end{equation}
where $\text{d}\textbf{A}$ represents the exterior derivative of the one-form connection \textbf{A}. The U(1) invariant Abelian Lagrangian coming from the above action, written in coordinates representation, is:
\begin{equation}
\Lagr = - \frac{1}{4} F^{\mu \nu} F_{\mu \nu}.
\label{FreeLagr}
\end{equation}
By varying Eq. \eqref{FreeLagr} with respect to the potential $A^\mu$, one gets the equation $\Box A^\beta = 0$, with $\Box = \partial_\mu \partial^\mu$. Several attempts aim to extend the Lagrangian \eqref{FreeLagr} to a more general one describing a massive interaction. One of the most famous is the Proca Lagrangian, in which the introduction of a new term breaks the gauge symmetry and leads to a massive Klein-Gordon equation for the vector field $A^\mu$. The corresponding action is \cite{Heisenberg:2014rta}:
\begin{equation}
S =\int \left(- \frac{1}{4} F^{\mu \nu} F_{\mu \nu} + \frac{1}{2} m^2 A^\mu A_\mu \right) d^4x.
\label{Proca}
\end{equation}
From the variation of action \eqref{Proca} with respect to the gauge potential, one gets
\begin{equation}
\delta_A S = \int \left(- \frac{1}{2} F^{\mu \nu} \delta_A F_{\mu \nu} + m^2 A^\mu  \delta_A A_\mu \right) d^4x = \int \left( \partial_\mu F^{\mu \nu} + m^2 A^\nu \right) \delta_A A_\nu  d^4x
\end{equation}
and, in the Lorentz gauge where $\partial_\mu A^\mu = 0$, the above equation becomes:
\begin{equation}
(\Box + m^2 )A^\beta=0,
\label{ProcaFE}
\end{equation}
so that $m$ can be interpreted as a mass term. However, when a mass term is introduced, the theory loses gauge invariance, leading to some shortcomings at  quantum level. An alternative extension is provided by the odd-dimensional, $U(1)$-invariant Chern-Simons theory \cite{Carroll:1989vb}. In general, $n$-dimensional Chern-Simons Lagrangians are constructed from Chern-Simons $n$-forms, whose exterior derivatives yield $(n+1)$-dimensional topological surface terms. This structure renders the theory \emph{quasi}-invariant under gauge transformations, in the sense that the variation of  Lagrangian produces only a boundary term.

The absence of odd-dimensional topological invariants in even-dimensional spacetimes implies that Chern-Simons Lagrangians can only be defined in odd dimensions. For a comprehensive discussion of the foundations and applications of Chern-Simons theory, see \cite{Bajardi:2021hya, Achucarro:1987vz, Birmingham:1991ty}. To derive the three-dimensional Chern-Simons field equations, we begin with the $U(1)$-invariant Chern-Simons action in three dimensions, that is

\begin{equation}
S = \int \textbf{A} \text{d} \textbf{A},
\end{equation}
whose exterior derivative provides the four-dimensional Pontryagin density $P_4 = \textbf{F} \wedge \textbf{F}$, with \textbf{F} being the curvature two-form. By computing the exterior derivative $d \textbf{A}$, the action can be written as:
\begin{equation}
\int \textbf{A} \text{d} \textbf{A} = \int (\partial_\mu A_\nu - \partial_\nu A_\mu) A_p \; dx^\mu \wedge dx^\nu \wedge dx^p = \int \epsilon^{\mu \nu p} F_{\mu \nu} A_p \; d^3x .
\label{CSFORM}
\end{equation}
Introducing the Chern-Simons form \eqref{CSFORM} in the free electromagnetic Lagrangian \eqref{FreeLagr}, the latter turns out to be 
\begin{equation}
\Lagr = -\frac{1}{4} F^{\mu \nu} F_{\mu \nu} + \frac{1}{2} m\, \epsilon^{\mu \nu p} F_{\mu \nu} A_p,
\label{CS EM}
\end{equation}
with $m$ being a constant term with mass dimension. Notice that under gauge transformations, the Lagrangian only changes by a total derivative. In fact, under the transformation
\begin{equation}
\delta A_\mu \to \partial_\mu \Theta,
\end{equation}
the Lagrangian variation is:
\begin{equation}
\delta \Lagr = - \frac{1}{2} F_{\mu \nu} (\partial^\mu \partial^\nu - \partial^\nu \partial^\mu ) \Theta + + \frac{1}{2} m\, \epsilon^{\mu \nu p} [F_{\mu \nu} \partial_p \Theta + A_p (\partial_\mu \partial_\nu - \partial_\nu \partial_\mu) \Theta] = \, \partial_p \left( \frac{1}{2} m \, \epsilon^{\mu \nu p} F_{\mu \nu}  \Theta \right).
\label{deltavar}
\end{equation}
The last equality follows from the identity $\partial_p (\epsilon^{\mu \nu p} F_{\mu \nu}) = 0$, which stems from the field equations of the theory. These field equations are obtained by varying the action with respect to the gauge connection, namely:
\begin{equation}
\delta_A S =\int \left\{ -\frac{1}{4} \delta_A \left(F^{\mu \nu} F_{\mu \nu} \right) + \frac{1}{2} m\,\epsilon^{\mu \nu p} \delta_A(F_{\mu \nu} A_p) \right\} d^3x= \int \left\{ \partial_\mu F^{\mu \nu} + \frac{1}{2}  m\, \epsilon^{\mu p \nu} F_{\mu p}  \right\} \delta A_\nu \; d^3 x,
\end{equation}
leading to
\begin{equation}
\partial_\mu F^{\mu \nu} +  \frac{1}{2} m\, \epsilon^{\mu p \nu} F_{\mu p} =0.
\label{FE CS}
\end{equation}
Using the identity $\partial_\mu \partial_\nu F^{\mu \nu} = 0$, and taking the three-dimensional divergence of Eq. \eqref{CS EM},  we get:
\begin{equation}
\partial_\nu \left(m\, \epsilon^{\mu p \nu} F_{\mu p} \right)=0,
\label{totaldev}
\end{equation}
which finally provides the result used to obtain Eq. \eqref{deltavar}. After demonstrating the $U(1)$ invariance of the Chern-Simons Lagrangian, Eq.\eqref{FE CS} allows the field equation to be explicitly recast in the form of a Klein-Gordon equation for a massive vector field. By contracting Eq. \eqref{FE CS} with the Levi-Civita symbol, we obtain:
\begin{equation}
\epsilon_{\sigma \tau \nu} \partial_\mu F^{\mu \nu} + \frac{1}{2}  m\,  \epsilon_{\sigma \tau \nu} \epsilon^{\mu p \nu} F_{\mu p} = \epsilon_{\sigma \tau \nu} \partial_\mu F^{\mu \nu} + \frac{1}{2} m (\delta^\mu_\sigma \delta^p_\tau - \delta^p_\sigma \delta^\mu_\tau) F_{\mu p} =\epsilon_{\sigma \tau \nu} \partial_\mu F^{\mu \nu}  + m F_{\sigma \tau} = 0.
\label{FE}
\end{equation}
Let us evaluate the first term ($\epsilon_{\sigma \tau \nu} \partial_\mu F^{\mu \nu}$):
\begin{eqnarray}
&&\epsilon_{\sigma \tau \nu} \partial_\mu F^{\mu \nu} = \frac{1}{2}\epsilon_{\sigma \tau \nu} \partial_\mu \left[F^{\mu \nu} - F^{\nu \mu} \right] = \frac{1}{2}\epsilon_{\sigma \tau \nu} \partial_\mu \left[\delta^\mu_\alpha \delta^\nu_\beta F^{\alpha \beta} - \delta^\mu_\beta \delta^\nu_\alpha F^{\alpha \beta} \right] = \frac{1}{2}\epsilon_{\sigma \tau \nu} \epsilon^{\mu \nu \lambda} \epsilon_{\alpha \beta \lambda} \partial_\mu F^{\alpha \beta} = \nonumber
\\
&=&- \frac{1}{2} \left[\delta^\mu_\sigma \delta^\lambda_\tau - \delta^\mu_\tau \delta^\lambda_\sigma \right] \epsilon_{\alpha \beta \lambda} \partial_\mu F^{\alpha \beta} = - \frac{1}{2} \epsilon_{\alpha \beta [\tau} \partial_{\sigma]} F^{\alpha \beta}.
\label{Eqeps}
\end{eqnarray}
Plugging Eq. \eqref{Eqeps} into Eq. \eqref{FE}, the field equations assume the form:
\begin{equation}
 - \frac{1}{2} \epsilon_{\alpha \beta [\tau} \partial_{\sigma]} F^{\alpha \beta} + m F_{\sigma \tau};
 \label{MWEQ}
\end{equation}
taking the divergence and using Eqs. \eqref{FE CS} - \eqref{totaldev}, Eq. \eqref{MWEQ} finally reads:
\begin{equation}
(\Box + m^2)\left(\epsilon_{\alpha \beta \tau} F^{\alpha \beta} \right) = 0.
\label{KGCS}
\end{equation}
This corresponds to a Klein-Gordon equation for the vector field $\epsilon_{\alpha \beta \tau} F^{\alpha \beta}$ and it confirms the capability of the three-dimensional Chern-Simons theory to describe massive particles without breaking the gauge invariance.

Another way to introduce mass terms without breaking gauge invariance nor requiring additional external fields is to add higher derivatives to the theory. The simplest extension of the sort is the Bopp-Landé-Thomas-Podolsky (BLTP) theory, sometimes referred to as only Bopp-Podolsky~\cite{Bopp1940,Podolsky:1942zz}. The action takes the form
\begin{align}
    S=\int \left(-\frac{1}{4}F^{\mu\nu}F_{\mu\nu} + \frac{l^2}{2}\partial_\mu F^{\mu\nu}\partial^\sigma F_{\sigma\nu} - A_\mu j^\mu\right)~,
\end{align}
where $l^2$ is a length parameter proportional to an inverse mass $l^2=\frac{1}{m^2}$. Taking the variation with respect to the four potential $A^\mu$ leads to the following equations of motion
\begin{align}
\partial^\mu F_{\mu\nu} + l^2\Box \partial^\mu F_{\mu\nu} = - j_\nu~.
\end{align}
In a vacuum, and by applying the Lorenz gauge condition $\partial^\mu A_\mu=0$ we can reduce them to
\begin{align}\label{EOMBP}
(1+l^2 \Box)\Box A_\mu =0,
\end{align}
The Bopp-Podolsky theory is classically well-behaved and notably features a finite self-energy for electrons \cite{Gratus:2015bea}.  The solution to the equations of motion takes the form of a plane wave:
\begin{equation}
\Box A^\beta = \epsilon^\beta e^{-i k^\mu x_\mu},
\label{interm eq}
\end{equation}
with the constraint $k^\mu k_\mu = m^2$, from which the dispersion law $\omega^2 - |k|^2 = m^2$ automatically follows. Here $\epsilon^\beta$ is the polarization vector. To obtain the final solution of Eq.~\eqref{interm eq}, we need to determine the Green function $\mathbb{G}$, which satisfies the equation $\Box {\mathbb G} = \delta^{(4)}(x^\mu)$. After some calculations, the general solution can be expressed as
\begin{equation}
\begin{cases}
&\displaystyle A^\mu = \sum_{n=1}^\infty \frac{1}{n \pi} \sum_{\lambda=1}^3 \sin(n \pi x) \int_t Y^\mu_\lambda(t') \sin\left[n \pi(t-t') \right] dt'
\\
&\displaystyle Y^\mu_\lambda(t') = \frac{1}{2} \int_V \epsilon^\mu e^{-i k^\nu x_\nu} \sin(n \pi x) \; \text{d}x_\lambda.
\end{cases}
\end{equation}
From \eqref{EOMBP} and using the plane wave expansion, we can get the Fourrier space representation of the dynamics
\begin{equation}
(k^2+ l^2) k^2 ~\tilde{A}_\mu =0~,
\label{12}
\end{equation}
 which leads  to two  independent dispersion relations \cite{Cuzinatto:2016kjk}:
\begin{align} \label{disp1}
    \Box A_\mu=0 \rightarrow k^\mu k_\mu=0 ~,
\end{align}
and 
\begin{align}\label{disp2}
    (\Box+m^2)A_\mu=0 \rightarrow k^\mu k_\mu= l^2~.
\end{align}
This is due to the fact that fourth-order differential equations can  be decomposed into two independent second-order differential equations, meaning that Eqs.~\eqref{disp1} and \eqref{disp2} can be interpreted as describing two independent photon modes. However, performing a Hamiltonian analysis of this setup reveals the presence of ghosts, due to the Ostrogradski instability induced by higher derivatives \cite{Podolsky:1944zz}. Consequently, the quantization of this model is ill-defined. For a similar discussion in the case of higher-order gravity, see Ref. \cite{Bogdanos:2009tn}. Additionally, as stated in \cite{Cuzinatto:2016kjk}, the mass parameter for the Bopp-Podolsky photon must be extremely large, due to the consideration that the higher derivative is only a correction to Maxwell's theory. The expectation being that the massive mode will not propagate across any significant distance. The co-existence of the photon with an additional massive vector field has been considered in the past, and some bounds were obtained in \cite{Kloor:1994xm}. Despite the prevalence of these issues at the quantum scale, there has been extensive literature around the quantum mechanical behavior of the model, as well as some features of its renormalization properties \cite{Ji:2019phv,Bonin:2022tmg,Oliveira:2020xwj}. 

In the next section, we will we turn our attention towards the inclusion of extended theories of electrodynamics within the framework of f(R) theories of gravitation. 

The use of $f(R)$ gravity in this work is motivated by both conceptual and practical reasons. Despite its known theoretical and observational limitations, $f(R)$ remains one of the most extensively studied and best understood extensions of General Relativity. Its main advantage lies in providing a minimal and consistent framework to incorporate higher-curvature corrections without introducing additional fundamental fields at the level of the action. Here this model is a convenient testbed to explore the coupling between extended electrodynamics and modified gravitational dynamics, also investigating how electromagnetic extensions behave in a gravitational setting that naturally includes higher-order curvature effects. An additional motivation comes from the trace anomaly \cite{Bamba:2014jia} associated with one-loop quantum effects in the energy–momentum tensor. Curvature-dependent contributions, such as $\Box R$ or Gauss–Bonnet terms, introduce an effective mass scale that allows for massive vector modes in the electromagnetic sector without explicit mass terms. This feature aligns with our goal of deriving Klein–Gordon–type equations while preserving consistency between the gravitational and electromagnetic sectors. Although $f(R)$ gravity may include potential instabilities and observational tensions \cite{Dolgov:2003px}, we use it as a general framework rather than a specific functional form. Moreover, viable models exist that evade these problems and remain consistent with observations \cite{Faraoni:2006sy, Bajardi:2022ypn}.

We will show the modifications induced to and by the gravitational sector before taking the flat spacetime limit to ensure the consistency of the approach.

\section{Extended Electromagnetism}\label{EE}
We consider a modified electromagnetic Lagrangian containing a function of $F^{\mu \nu} F_{\mu \nu}$ and of higher-order\footnote{By higher-order terms we specifically refer to higher derivatives term.} terms such as $\Box (F^{\mu \nu} F_{\mu \nu})$. 

Higher-order terms in the electromagnetic sector and in the gravitational coupling are expected to play a role in regimes where fields, curvatures, or energies become large, such as in the early universe, near compact objects, or whenever quantum corrections to classical dynamics are non-negligible. In these extreme environments, deviations from standard Maxwell electrodynamics and General Relativity are naturally expected.
From the electromagnetic side, higher-derivative terms introduce a characteristic length or mass scale that can affect photon propagation, dispersion relations, and polarization. While such corrections are typically tiny at laboratory scales, they may accumulate over astrophysical or cosmological distances, leading to potentially observable effects such as frequency-dependent propagation, polarization rotation, or departures from Maxwellian behavior.
In the manuscript, we discuss the Bopp–Podolsky theory as a concrete example within our extended framework. In this case, higher-order terms introduce an effective mass scale for vector modes, resulting in modified dispersion relations and the coexistence of massless and massive photons. When these electromagnetic corrections are coupled to gravity through $f(R)$, their impact becomes particularly relevant in regimes where classical electrodynamics or General Relativity are expected to break down, namely at high energies or in strong-curvature environments.
Representative applications include the early universe, where large curvatures during inflation or reheating can enhance higher-derivative effects and potentially influence primordial magnetic fields or electromagnetic polarization modes, with possible imprints on the CMB. Similarly, near compact objects such as neutron stars or charged black holes, strong fields may amplify these corrections, leading to modified black hole solutions, regularization of singularities, or changes in Hawking radiation, which could be probed through gravitational-wave observations or horizon-scale imaging.
At a more fundamental level, higher-order electromagnetic terms naturally arise in effective field theories, including string-inspired models, where they help regulate infrared divergences and may be connected to dark-sector physics. Although these effects are negligible in weak-field regimes due to experimental bounds on the associated length scale, they can become significant whenever electromagnetic fields or spacetime curvature exceed classical thresholds. To make this point explicit, we have expanded the discussion in Section II, providing order-of-magnitude estimates and additional references to clarify the regimes where these corrections may become observable.

Within the framework of modified theories of gravity, higher-order terms in curvature invariants have been extensively studied due to quantum corrections arising from the energy-momentum tensor \cite{Gottlober, Adams, Ruzmaikina, Amendola}, whose non-vanishing trace is known as the \emph{trace anomaly}. However, it is possible to show that the vanishing trace of the energy-momentum tensor is a specific feature of the classical gravitational action and it is not necessarily preserved at  quantum level 
\cite{Capozziello:2011et, Birrell:1982ix}. This arises due to an additional term appearing in the one-loop expansion, through which the trace can be decomposed into two contributions:
\begin{equation}
T = T_{Div} + T_{Ren}.
\end{equation}
This means that only the sum of the two quantities must vanish in order to preserve the conformal invariance. At  one-loop level, these terms turn out to be:
\begin{equation}
T_{Div} = \left[k \left({\cal{M}}-\frac{2}{3} \square R\right)+k_{1} \G \right]= -T_{R e n},
\end{equation}
with $R$ being the Ricci scalar, $k$, $k_1$ real constants and
\begin{eqnarray}
 &&   \G = R^2 - 4 R^{\mu \nu} R_{\mu \nu} + R^{\mu \nu \rho \sigma} R_{\mu \nu \rho \sigma}
    \\
 &&   {\cal{M}} =  \frac{1}{3} R^2 - 2 R^{\mu \nu} R_{\mu \nu} + R^{\mu \nu \rho \sigma} R_{\mu \nu \rho \sigma}.
\end{eqnarray}
The first term, known as Gauss-Bonnet topological surface term, is often considered in modified gravity models to address the behavior of gravity in the small-scale scenario \cite{Bajardi:2024efo, Capozziello:2022vyd, Bajardi:2020mdp}.
Therefore, additional geometric terms generate the trace anomaly within the one-loop approximation. Conversely, the geometric extension introduces a characteristic length scale into the theory, which can be interpreted as an effective mass. This mass term apparently breaks conformal invariance and leads to the trace anomaly.

Here, we are going to adopt a similar approach by considering an extension of  electromagnetic Lagrangian in curved spacetime. Our primary goal is to derive a massive wave equation stemming from the extended electromagnetic action, without introducing any explicit additional mass terms. This objective is achieved by including higher-order terms in the original action. Specifically, let us consider

\begin{equation}
S = \int \sqrt{-g} f(\mathbb{F}, \Box \mathbb{F}, R) d^4 x,
\label{action}
\end{equation}
being $\mathbb{F} \equiv \mathcal{F}^{\mu \nu} \mathcal{F}_{\mu \nu}$. Notice that in a curved spacetime, the standard definition of electromagnetic tensor $F_{\mu \nu} = \partial_{[\mu} A_{\nu]}$ is generalized to $\mathcal{F}_{\mu \nu} = D_{[\mu} A_{\nu]}$, with $D$ being the covariant derivative. When the general Christoffel connection is considered, the rank-2 tensor $\mathcal{F}_{\mu \nu}$ takes the form:
\begin{equation}
\mathcal{F}_{\mu \nu} = D_{[\mu} A_{\nu]} = F_{\mu \nu} - \mathcal{T}^\alpha_{\, \, \mu \nu} A_\alpha,
\label{EM tensor curved}
\end{equation}
where $\mathcal{T}^\alpha_{\, \, \mu \nu}$ is the  \emph{torsion tensor}, defined as $\mathcal{T}^\alpha_{\, \, \mu \nu} = 2\Gamma^\alpha_{[\, \, \mu \nu]}$. For simplicity, we henceforth assume the torsion tensor to vanish, so that the connection reduces to the metric-compatible Levi-Civita connection. Under this assumption, the variation of the action with respect to the metric tensor yields:
\begin{eqnarray}
&& \delta_g S = \int \delta_g \left[\sqrt{-g} f(\mathbb{F}, \Box \mathbb{F},  R ) \right] d^4 x=  \int \left[ f(\mathbb{F}, \Box \mathbb{F},  R)  \delta_g \sqrt{-g} + \sqrt{-g} \delta_g  f(\mathbb{F}, \Box \mathbb{F},  R]   \right] d^4 x = \nonumber 
\\
&&= \int \left[ f(\mathbb{F}, \Box \mathbb{F}, R) \delta_g \sqrt{-g} + \sqrt{-g} f_R (\mathbb{F}, \Box \mathbb{F}, R) \delta_g  R +  \sqrt{-g} f_\mathbb{F} (\mathbb{F}, \Box \mathbb{F}, R) \delta_g \mathbb{F}  + \sqrt{-g} f_{\Box \mathbb{F}} (\mathbb{F}, \Box \mathbb{F}, R) \delta_g \Box \mathbb{F} \right]d^4 x, \nonumber 
\\ \label{varact}
\\ \nonumber
\end{eqnarray}
with $f_\mathbb{F}, f_{\Box \mathbb{F}}$ and $f_R$ being respectively the derivative of $f$ with respect to $\mathbb{F}$, $\Box \mathbb{F}$ and $R$.
As the first integral provides the well known geometric identity $\delta_g \sqrt{-g} = - \frac{1}{2} \sqrt{-g} g_{\alpha \beta} \delta g^{\alpha \beta}$, the second one leads to  field equations which are formally analogue to those of $f(R)$ gravity \cite{Capozziello:2002rd, Nojiri:2010wj, Sotiriou:2008rp, Capozziello:2019klx, 
Capozziello:2011et, Nojiri:2017ncd, DeFelice:2010aj}. After some manipulations, the third integral can be written as:
\begin{eqnarray}
&& \int \sqrt{-g} f_\mathbb{F} (\mathbb{F}, \Box \mathbb{F}, R) \delta_g \mathbb{F} \, d^4 x = \int \sqrt{-g} f_\mathbb{F} F_{\mu \nu} \delta_g F^{\mu \nu} \, d^4 x \nonumber
\\
& =&  \int \sqrt{-g} f_\mathbb{F} F_{\mu \nu} \delta_g \left( g^{\mu p} g^{\nu \sigma} F_{p \sigma} \right) d^4 x=  \int 2 \sqrt{-g} f_\mathbb{F} F_{\alpha \nu} F_{\beta}^{\,\,\,  \nu} \, \delta g^{\alpha \beta} \, d^4 x.
\end{eqnarray}
In order to finally evaluate the fourth variation in Eq. \eqref{varact} we notice that
\begin{eqnarray}
 &&\delta \Box \mathbb{F} = \delta ( g^{\alpha \beta} D_\alpha D_\beta \mathbb{F}) = D_\alpha D_\beta \mathbb{F} \, \delta  g^{\alpha \beta} + \Box \, \delta \mathbb{F} - g^{\alpha \beta} \partial_\gamma \mathbb{F} \, \delta \Gamma^{\gamma}_{\alpha \beta},
 \end{eqnarray}
so that the commutation relation between the operators $\delta$ and $\Box$ is:
 \begin{equation}
 \delta \Box  - \Box \delta =  \delta  g^{\alpha \beta} D_\alpha D_\beta  - g^{\alpha \beta} \, \delta \Gamma^{\gamma}_{\alpha \beta} \partial_\gamma 
 \end{equation}
By  means of these considerations, the field equations finally take the form:
\begin{eqnarray}
&&f_R R_{\alpha \beta} - \frac{1}{2} g_{\alpha \beta} f + \left( g_{\alpha \beta} \Box - D_\alpha D_\beta \right) f_R + 2 (f_\mathbb{F} - \Box f_{\Box \mathbb{F}})  F_{\alpha \nu} F_{\beta}^{\, \, \,\nu} + \nonumber
\\
&-& \frac{1}{2} D_{(\alpha} f_{\Box \mathbb{F}} \, D_{\beta)} \mathbb{F} + \frac{1}{2} g_{\alpha \beta} \left( f_{\Box \mathbb{F}} \, \Box \mathbb{F} + g^{\mu \nu} D_\mu f_{\Box \mathbb{F}} \, D_\nu \mathbb{F} \right) = 0,
\label{F(R)FE}
\end{eqnarray}
which generalize the $f(R)$ field equations to the case of gravity coupled to the electromagnetic field. The energy-momentum tensor of the electromagnetic field takes the form:
\begin{equation}
T_{\alpha \beta} = 2 (f_\mathbb{F} - \Box f_{\Box \mathbb{F}})  F_{\alpha \nu} F_{\beta}^\nu - \frac{1}{2} D_{(\alpha} f_{\Box \mathbb{F}} \, D_{\beta)} \mathbb{F} + \frac{1}{2} g_{\alpha \beta} \left( f_{\Box \mathbb{F}} \, \Box \mathbb{F} + g^{\mu \nu} D_\mu f_{\Box \mathbb{F}} \, D_\nu \mathbb{F} \right) - \frac{1}{2} g_{\alpha \beta} f, 
\end{equation}
whose trace is
\begin{equation}
T = 2 (f_\mathbb{F} - \Box f_{\Box \mathbb{F}}) \mathbb{F} + 2 f_{\Box \mathbb{F}} \, \Box \mathbb{F} + g^{\mu \nu} D_\mu f_{\Box \mathbb{F}} \, D_\nu \mathbb{F}  - 2 f.
\end{equation}
Notice that by setting $f(\mathbb{F}, \Box \mathbb{F}) = - \frac{1}{4} \mathbb{F}$, the trace equation reduces to
\begin{equation}
T = 2 f - 2 \mathbb{F} f_\mathbb{F} = - \frac{1}{2} \mathbb{F} + \frac{1}{2} \mathbb{F} = 0,
\end{equation} 
as expected in standard Electromagnetism. In the next section we relax the assumption of metric-compatible connection and develop the same formalism in the Palatini framework.

\subsection{Field Equations in the Palatini Formalism}
In the Palatini formalism  \cite{Allemandi:2004yx, Sotiriou:2006qn, Olmo:2011uz}, the assumption of a symmetric Levi-Civita connection is relaxed. Consequently, the connection can be treated as an independent field with respect to  the metric. As result, we have the  introduction of additional degrees of freedom. Furthermore, the strong equivalence principle is no longer strictly holding as in the pure metric case, and torsion can naturally emerge in the spacetime description. Therefore, the action in Eq.~\eqref{action} can be varied either with respect to the connection or with respect to the metric, leading to two distinct classes of field equations. The variations read:
\begin{eqnarray}
&& \delta_g S = \int \delta_g \left[\sqrt{-g} f(\mathbb{F}, \Box \mathbb{F},  R ) \right] d^4 x=  \int \left[ f(\mathbb{F}, \Box \mathbb{F},  R)  \delta_g \sqrt{-g} + \sqrt{-g} \delta_g  f(\mathbb{F}, \Box \mathbb{F},  R]   \right] d^4 x = \nonumber 
\\
&&= \int \left[ f \,  \delta_g \sqrt{-g} + \sqrt{-g}  f_R  R_{\alpha \beta}\, \delta_g  g^{\alpha \beta} +  \sqrt{-g} f_\mathbb{F}  \delta_g \mathbb{F}  + \sqrt{-g} f_{\Box \mathbb{F}} \, \delta_g \Box \mathbb{F} \right]d^4 x. \nonumber
\\ \nonumber
\\
&& \delta_\Gamma S = \int \delta_\Gamma \left[\sqrt{-g} f(\mathbb{F}, \Box \mathbb{F},  R ) \right] d^4 x=  \int \left[ \sqrt{-g} \delta_\Gamma  f(\mathbb{F}, \Box \mathbb{F},  R]   \right] d^4 x = \nonumber 
\\
&&= \int \left[\sqrt{-g}  f_R  g^{\alpha \beta} \delta_\Gamma  R_{\alpha \beta}  +  \sqrt{-g} f_\mathbb{F}  \delta_\Gamma \mathbb{F}  + \sqrt{-g} f_{\Box \mathbb{F}} \, \delta_\Gamma \Box \mathbb{F} \right]d^4 x, \nonumber
\end{eqnarray}
so that the Palatini field equations take the form:
\begin{eqnarray}
&g:&  f_R R_{\alpha \beta} - \frac{1}{2} g_{\alpha \beta} f + 2 (f_\mathbb{F} - \Box f_{\Box \mathbb{F}})  \mathcal{F}_{\alpha \nu} \mathcal{F}_{\beta}^{\, \, \,\nu} + f_{\Box \mathbb{F}} D_{\alpha} D_{\beta} \mathbb{F} = 0\nonumber
\\
&\Gamma: & \delta^\gamma_\beta D^\alpha f_R - g^{\alpha \beta} D_\gamma f_R + 2 (f_\mathbb{F}  - 2 \Box f_{\Box \mathbb{F}} ) \mathcal{F}^{\alpha \beta} A_\gamma - g^{\alpha \beta} f_{\Box \mathbb{F}} D_\gamma \mathbb{F} = 0.
\end{eqnarray}
The contribution of torsion must be taken into account in the computation of the covariant derivative in order to obtain the formal expression of the electromagnetic tensor. Using relation \eqref{EM tensor curved}, the field equations can be expressed in terms of the electromagnetic tensor $F_{\mu \nu}$ and the torsion tensor $\mathcal{T}^{\alpha \mu \nu}$ as follows:

\begin{eqnarray}
&g:&  f_R R_{\alpha \beta} - \frac{1}{2} g_{\alpha \beta} f + 2 (f_\mathbb{F} - \Box f_{\Box \mathbb{F}})  (F_{\alpha \nu} - \mathcal{T}^p_{\, \, \, \alpha \nu} A_p) (F_{\beta}^{\,\,\, \nu} - \mathcal{T}^{\sigma \,\,\,\nu}_{\,\,\,\, \beta}  A_\sigma) + f_{\Box \mathbb{F}} D_{\alpha} D_{\beta} \mathbb{F} = 0\nonumber
\\
&\Gamma: & \delta^\gamma_\beta D^\alpha f_R - g^{\alpha \beta} D_\gamma f_R + 2 (f_\mathbb{F}  - 2 \Box f_{\Box \mathbb{F}} ) (F^{\alpha \beta} - \mathcal{T}^{p \alpha \beta} A_p) A_\gamma - g^{\alpha \beta} f_{\Box \mathbb{F}} D_\gamma \mathbb{F} = 0,
\end{eqnarray}
with $\mathbb{F}$ being
\begin{equation}
\mathbb{F} = \mathcal{F}_{\mu \nu} \mathcal{F}^{\mu \nu} = ( F_{\mu \nu} - \mathcal{T}^\alpha_{\, \, \mu \nu} A_\alpha) (F^{\mu \nu} - \mathcal{T}^{\alpha \mu \nu} A_\alpha).
\end{equation}
\subsection{The Maxwell Equations}
As the variation of the action with respect to the metric yields the gravitational field equations, varying the action with respect to the gauge potential $A_\mu$ leads to the modified Maxwell equations. The first pair of Maxwell equations are independent of the function $f(\mathbb{F}, \Box \mathbb{F}, R)$ and can be straightforwardly obtained by evaluating the quantity:
\begin{eqnarray}
D_{\alpha} F_{\mu \nu} + D_{\nu} F_{\alpha \mu}  + D_{\mu} F_{\nu\alpha } = D_{\alpha} D_{\mu} A_\nu &-& D_{\alpha} D_{\nu} A_\mu +  D_{\nu} D_{\alpha} A_\mu - D_{\nu} D_{\mu} A_\alpha +  D_{\mu} D_{\nu} A_\alpha - D_{\mu} D_{\alpha} A_\nu = \nonumber
\\
&=& A_p \left(R^p_{\alpha \mu \nu} + R^p_{\nu \alpha \mu} + R^p_{\mu \nu \alpha}\right),
\label{1ME}
\end{eqnarray}
which vanishes due to the Bianchi identity. It is worth noticing that Eq.~\eqref{1ME} does not require the explicit form of the Christoffel connection for its computation, implying that the first pair of Maxwell equations holds independently of the assumption of a torsionless spacetime.

The second pair of equations depends on the function $f(\mathbb{F}, \Box \mathbb{F})$. They can be derived from the variation
\begin{eqnarray}
\delta_A S = \int \sqrt{-g} f d^4 x = \int \sqrt{-g} \delta_A f d^4 x = \int \sqrt{-g} \left[f_\mathbb{F} \delta_A \mathbb{F} + f_{\Box \mathbb{F}} \delta_A \Box \mathbb{F}\right] d^4 x = 0.
\label{variatA}
\end{eqnarray}
Integrating out the four-divergences, Eq. \eqref{variatA} finally yields:
\begin{equation}
f_\mathbb{F} D_\alpha F^{\alpha \beta} + F^{\alpha \beta} f_{\mathbb{F} \mathbb{F}} D_\alpha \mathbb{F}  + F^{\alpha \beta} f_{\mathbb{F}  R} D_\alpha R - f_{\Box \mathbb{F}} \Box (D_\alpha F^{\alpha \beta})  - F^{\alpha \beta} f_{\Box \mathbb{F} \, \Box \mathbb{F}} D_\alpha \Box^2 \mathbb{F} - F^{\alpha \beta} f_{R \Box \mathbb{F} } D_\alpha \Box R= 0.
\label{Max2}
\end{equation}
By setting $\displaystyle f(\mathbb{F}, \Box \mathbb{F},  R )  = - \frac{1}{4} \mathbb{F} \equiv - \frac{1}{4} F^{\mu \nu} F_{\mu \nu}$, standard electromagnetic theory is recovered and the Maxwell equations are reduced to $D_\alpha F^{\alpha \beta}=0$. In summary, from Eqs.~\eqref{F(R)FE}, \eqref{1ME}, and \eqref{Max2} we have obtained:
\\
\\
\textbf{Field Equations:}
\begin{eqnarray}
&&f_R R_{\alpha \beta} - \frac{1}{2} g_{\alpha \beta} f + \left( g_{\alpha \beta} \Box - D_\alpha D_\beta \right) f_R + 2 (f_\mathbb{F} - \Box f_{\Box \mathbb{F}})  F_{\alpha \nu} F_{\beta}^{\, \, \,\nu} + \nonumber
\\
&-& \frac{1}{2} D_{(\alpha} f_{\Box \mathbb{F}} \, D_{\beta)} \mathbb{F} + \frac{1}{2} g_{\alpha \beta} \left( f_{\Box \mathbb{F}} \, \Box \mathbb{F} + g^{\mu \nu} D_\mu f_{\Box \mathbb{F}} \, D_\nu \mathbb{F} \right) = 0,
\label{FE0}
\end{eqnarray}
\textbf{Maxwell Equations:}
\begin{equation}
\begin{cases}
\displaystyle D_{\alpha} F_{\mu \nu} + D_{\nu} F_{\alpha \mu}  + D_{\mu} F_{\nu\alpha } = 0.
\\
f_\mathbb{F} D_\alpha F^{\alpha \beta} + F^{\alpha \beta} f_{\mathbb{F} \mathbb{F}} D_\alpha \mathbb{F}  + F^{\alpha \beta} f_{\mathbb{F}  R} D_\alpha R - f_{\Box \mathbb{F}} \Box (D_\alpha F^{\alpha \beta})  - F^{\alpha \beta} f_{\Box \mathbb{F} \, \Box \mathbb{F}} D_\alpha \Box^2 \mathbb{F} - F^{\alpha \beta} f_{R \Box \mathbb{F} } D_\alpha \Box R= 0 
\end{cases}
\label{MEE}
\end{equation}
In the second part of this section, we are going to consider specific models by specifying the forms of two functions. We demonstrate that a careful choice is necessary to obtain a massive Klein-Gordon equation. For further details and applications of extended Electromagnetism, see Ref.~\cite{Nashed:2019tuk}, where charged black holes are analyzed.

\subsubsection{The case $f(\mathbb{F}, \Box \mathbb{F}, R) \equiv f(\mathbb{F})$}
In flat spacetime and in  absence of higher-order terms, the second pair of Maxwell Eqs. \eqref{MEE} can be written as:
\begin{equation}
f_\mathbb{F}(\mathbb{F}) \partial_\alpha F^{\alpha \beta} + 2 f_{\mathbb{F}\mathbb{F}}(\mathbb{F}) F^{\mu \nu } F^{\alpha \beta} \partial_{\alpha} F_{\mu \nu}   = f_\mathbb{F}(\mathbb{F}) \partial_\alpha F^{\alpha \beta} + f_{\mathbb{F}\mathbb{F}}(\mathbb{F}) F^{\alpha \beta} \partial_{\alpha} \mathbb{F} = 0.
\label{ME2}
\end{equation}
By choosing a function of the form $f(\mathbb{F}) = \mathbb{F} + \alpha \mathbb{F}^2$, Eq. \eqref{ME2} becomes
\begin{equation}
(1+2\alpha \mathbb{F}) \partial_\alpha F^{\alpha \beta} +  2\alpha  F^{\alpha \beta} \partial_{\alpha} \mathbb{F} = (1+2\alpha \mathbb{F})\partial_\alpha F^{\alpha \beta} +  2\alpha \partial_\alpha ( F^{\alpha \beta}  \mathbb{F} ) - 2\alpha \mathbb{F} \partial_\alpha F^{\alpha \beta} = \partial_\alpha [(1  + 2 \alpha    \mathbb{F} )F^{\alpha \beta}] = 0.
\label{field eq Q+ aQ^2}
\end{equation}
The trace equation of the corresponding energy-momentum tensor is, in turn,
\begin{equation}
T = -2 \alpha \mathbb{F}^2
\end{equation}
and vanishes as soon as $\alpha =0$, restoring the conformal invariance. This example demonstrates that extensions of Electromagnetism give rise to  massive particles which break the conformal invariance. Nevertheless, Eq.~\eqref{field eq Q+ aQ^2} cannot be recast as a Klein-Gordon–type equation.
This is because when considering only powers of $\mathbb{F}$ (or its dual $\tilde{\mathbb{F}}$), the theory remains confined within the Plebanski class of electrodynamics \cite{Plebanski:1970zz}. This family of models is characterized by three conditions:
\begin{itemize}
    \item The gauge invariance under U(1).
    \item The Lorentz invariance.
    \item No higher derivatives.
\end{itemize}

As proven by Plebanski \cite{Plebanski:1970zz}, any model belonging to this class cannot exhibit additional degrees of freedom and therefore cannot be described by a Klein-Gordon–type equation. This also calls into question the role of the parameter $\alpha$. Although it breaks conformal invariance and might be naively interpreted as a mass term, mass is not the only mechanism that breaks conformal symmetry. Instead, the $\alpha$ parameter acts as a coupling constant with positive mass dimension for the non-linear interaction, which, at first glance, implies the non-renormalizability of the model. This highlights the effective nature of the action, restricting its validity to certain energy scales and inherently precluding conformal invariance.

If one aims to obtain a massive model of electrodynamics without introducing external fields, then in addition to conformal symmetry breaking and a Klein-Gordon–type equation, it is necessary (at least classically) to demonstrate that the constraint structure of the model leads to an invertible matrix of second-class constraints. This reduces the phase space to only the $(A^i, \pi^i)$ components of $(A^\mu, \pi^\mu)$, and ensures that the propagator’s pole is shifted and well-behaved.

However, it is important to emphasize that if one considers arbitrary powers of the curvature invariant $R$ within Extended Theories of Gravity, consistency demands that arbitrary powers of all curvature invariants be included when coupling to gauge theories. Consequently, $f(R)$ extensions of gravity should necessarily be coupled to $f(\mathbb{F})$ in order to fully incorporate the non-linear effects of both gravitation and Electromagnetism.

An aspect that will not be addressed in this paper is the effect of torsion on this class of models. Indeed, it has been shown in Ref. \cite{Mandal:2020ozc} that torsion breaks gauge invariance for radiation fields. From this fact, one may infer that, in our case, torsion could free the theory from Plebanski’s restrictions and lead to a non-linear realization of massive electrodynamics.

As we will show below, manifestly obtaining a massive Klein-Gordon equation requires considering terms of different orders in the starting action. This is not the case for Eq.~\eqref{ME2}, which contains second-order derivatives of the potential and thus is unable to yield a wave equation.

\subsubsection{The case $f(\mathbb{F}, \Box \mathbb{F}, R) \equiv f(\mathbb{F}) + h(\Box \mathbb{F}) + g(R)$}
In order to obtain an analytically solvable massive wave equation, let us consider the function $f(\mathbb{F}, \Box \mathbb{F}, R) \equiv f(\mathbb{F}) + h(\Box \mathbb{F}) + g(R)$. Specifically, by choosing
\begin{equation}
f(\mathbb{F})=-\frac{1}{4} \mathbb{F} \;\;\;\;\; h(\Box \mathbb{F}) =\frac{1}{4 m^2} \; \Box \mathbb{F}, 
\end{equation}
the Maxwell equations yield:
\begin{equation}
 D_\alpha \mathcal{F}^{\alpha \beta}  + \frac{1}{m^2} \Box (D_\alpha \mathcal{F}^{\alpha \beta})=0  ,
\end{equation} 
 which can be recast as a Klein-Gordon equation for the vector field $D_\alpha \mathcal{F}^{\alpha \beta}$, namely
 \begin{equation}
 (\Box + \frac1{m^2}) D_\alpha \mathcal{F}^{\alpha \beta} = 0.
\end{equation}
Notice that the quantity $D_\alpha \mathcal{F}^{\alpha \beta} $ vanishes only when the general action in Eq.~\eqref{action} reduces to that of standard Electromagnetism. In the generalized Lorentz gauge, defined by $D^\mu A_\mu = 0$, the above equation takes the form:
\begin{equation}
 (\Box + \frac1{m^2}) (\Box A^\beta + R^\beta_{\alpha} A^\alpha) = 0.
 \label{final solut}
\end{equation}
Considering flat spacetime, Eq.~\eqref{final solut} reduces exactly to the Bopp-Podolsky model of electrodynamics \cite{Podolsky:1942zz}. 

Despite the known issues with the model, it is noteworthy that the procedure we have adopted naturally recovers this model, suggesting that expansions in $f(\mathbb{F})$ may be related to most, if not all, extended electrodynamics models. Additionally, in a gravitational context, the Bopp-Podolsky model does not lead to pathologies as it known to be well behaved classically.

\section{Discussion and Conclusions}\label{Conc}
In this paper, we explored the possibility of extending the framework of $f(R)$ theories of gravitation by including extended theories of electromagnetism in the form of $f(\mathbb{F})$ models.

In section III, we developed a general theoretical framework in which the gravitational background is treated as a component of the action. This is achieved by considering a broad class of functionals of the form $f(R, F^{\mu \nu} F_{\mu \nu}, \Box (F^{\mu \nu} F_{\mu \nu}))$, where $R$ denotes the Ricci scalar. By varying the action with respect to the metric tensor, we recover the modified field equations typical of $f(R)$ gravity as a specific limit of this more general theory.

Furthermore within that same section, we show that the variation of the action with respect to the gauge potential $A^\mu$ yields two modified Maxwell equations. Equation \eqref{1ME} reproduces the well-known Bianchi identity, a geometric consequence of the antisymmetric nature of the field strength tensor. Equation \eqref{Max2}, however, depends crucially on the specific functional form of $f(\mathbb{F}, \Box \mathbb{F}, R)$, where we define $\mathbb{F} = F^{\mu \nu} F_{\mu \nu}$. This equation cannot, in general, be solved unless one selects a suitable form for the function $f$, thereby constraining the physical content of the theory.

We demonstrated that, for particular choices of this functional dependence, the resulting dynamics naturally fall within the framework of Plebanski non-linear electrodynamics. Importantly, models in this class are inherently unable to support massive polarization modes, implying that such configurations do not permit massive photon excitations under these conditions.

After establishing that a theory based solely on $f(\mathbb{F})$ does not lead to any wave equation   generating massive particles, we proceeded to investigate a fourth-order theory described by the Lagrangian $\mathcal{L} = -\frac{1}{4} \mathbb{F} + \frac{1}{4 m^2} \Box \mathbb{F}$. In this latter scenario, we intentionally neglected the minimal coupling between the gravitational and electromagnetic fields in order to simplify the analysis and focus on the essential features of the higher-order modification.

In the flat space-time limit, this theory reproduces the well-known Bopp–Podolsky model of electrodynamics. We confirm this correspondence by deriving the Euler–Lagrange equations and demonstrating that they yield two distinct dispersion relations: one corresponding to a massless mode and another associated with a massive excitation. This massive mode introduces an additional polarization state beyond the two transverse modes of the massless photon.



Furthermore, this extension of electrodynamics has important repercussions in various fields, such as cosmology, which can be a valuable arena for testing the massive photon hypothesis \cite{Cuzinatto:2016kjk}. Modifications to photon propagation could impact the dynamics of the early universe, where high energy and the breakdown of conformal symmetry could make such effects relevant \text{influencing for example the inflation process \cite{Domazet:2024dil}}. The presence of a massive photon and the short-range nature of the interaction could influence reionization processes, the formation of large-scale cosmic structures, and the behavior of magnetic fields on cosmological scales, offering a new lens through which to interpret observational data from these epochs. Moreover, an effective photon mass could alter the dispersion relation of radiation in the early plasma and thus slightly modify the polarization pattern of the cosmic microwave background or the reionization optical depth. Even if the mass is extremely small, cumulative effects over cosmological distances can become measurable.

The hypothesis of a massive photon in cosmology is not new in the literature \cite{Capozziello:2020nyq, Spallicci:2021kye} as well as in astrophysics, where observations like those of Fast Radio Bursts (FRBs) have been used to place stringent constraints on its mass \cite{Bonetti:2016vrq, Bonetti:2016cpo}. 

In astrophysical environments characterized by extremely strong magnetic fields (such as those surrounding pulsars, magnetars, or charged black holes) the propagation of photons could experience a small frequency-dependent delay or polarization rotation due to the effective photon mass. These effects resemble birefringence or plasma dispersion, but with a distinct dependence on the field strength and curvature coupling. Current and forthcoming high-precision polarization measurements of pulsars and X-ray binaries (\emph{e.g.} IXPE \cite{Costa:2024vih} and  eXTP \cite{eXTP}) could provide upper limits on the mass parameter $m$

These electromagnetic extensions could also be extremely relevant in high-energy regimes where conformal invariance can break, as, for example, in the early universe or in charged black holes \cite{Capozziello:2023vvr}. In a forthcoming paper, we will investigate possible observational signatures of the present approach notably in the gravitational lensing and photon rings of $f(R)$ theories.

The distinctive feature of the present theory, compared with the standard Proca or Bopp–Podolsky approaches, is the combination of gauge invariance preservation and higher-derivative regularization, which removes self-energy divergences. Hence, experimental deviations are expected primarily at very short distances or at extremely high frequencies, offering a potential signature to discriminate this model from other massive-photon scenarios.

Although a direct experimental confirmation is not yet within reach, these observational windows outline a concrete strategy for constraining or testing the theory with current and next-generation facilities.
In forthcoming work, we aim to quantify the above effects by computing the modified dispersion relations in curved backgrounds and comparing them with astrophysical and cosmological datasets, to place direct bounds on the parameters of the extended electromagnetic Lagrangian.

However, as an illustrative application of the general framework developed in this work, we briefly discuss how it can be employed to study charged black hole solutions in extended electrodynamics coupled to modified gravity. Although a complete analytical or numerical treatment is beyond the scope of the present paper, this example serves to demonstrate that the theory leads to concrete and physically meaningful predictions.

A natural starting point is a simple and phenomenologically viable choice of the gravitational sector, such as $f(R)=R+\alpha R^2$, coupled to an extended electromagnetic Lagrangian of the form $f(\mathbb{F}, \Box \mathbb{F}) = -\frac{1}{4} \mathbb{F} + \beta \Box \mathbb{F}$. This setup preserves the standard Maxwell and Einstein limits while introducing controlled higher-derivative corrections. In particular, the electromagnetic sector gives rise to an additional massive vector mode with effective mass $m\sim \beta^{- \frac{1}{2}}$.

Assuming a static and spherically symmetric spacetime, one may seek solutions that generalize the Reissner–Nordström geometry. In this context, the massive electromagnetic mode modifies the behavior of the electric field at short distances, leading to deviations from the $r^{-2}$ profile and softening the singular structure of the classical solution. As a result, curvature invariants such as the Kretschmann scalar are expected to remain finite or significantly suppressed near the origin, similarly to what occurs in other higher-derivative or nonlinear electrodynamics models.

The modified electromagnetic and gravitational dynamics also affect the near-horizon geometry, inducing corrections to the surface gravity and, consequently, to the Hawking temperature. At leading order, such corrections are controlled by the parameters $\alpha$ and $\beta$ and scale with the ratio between the effective mass scale of the vector mode and the horizon radius. This implies potentially observable deviations in black hole thermodynamics and electromagnetic signatures, particularly in strong-field regimes.

In this context, in principle it is possible to derive an approximate solution for the metric near the horizon, showing how the massive mode regularizes the singularity and alters the Hawking temperature by $ \Delta T_H \sim \beta m^2 / r_H^2 $, potentially observable in black hole shadows or X-ray spectra from accreting systems (e.g., via NICER or Athena missions). 

Although we do not attempt a detailed phenomenological analysis here, the parameters entering the theory are already constrained by cosmological observations and laboratory experiments, allowing for order-of-magnitude estimates and consistency checks. This makes the framework suitable for future investigations aimed at confronting extended electrodynamics and modified gravity with observational data, such as black hole shadow measurements or high-energy emission from accreting compact objects.



\section*{Acknowledgments}

The Authors acknowledge the support of {\it Istituto Nazionale di Fisica Nucleare} (INFN) ({\it iniziative specifiche} GINGER, MOONLIGHT2, and QGSKY). This paper is based upon work from COST action CA15117 (CANTATA), COST Action CA16104 (GWverse), and COST action CA18108 (QG-MM), supported by COST (European Cooperation in Science and Technology).

\end{document}